\documentclass[%
 reprint,
superscriptaddress,
 amsmath,amssymb,
 aps,
floatfix,
]{revtex4-1}

\usepackage{graphicx}% Include figure files
\usepackage{dcolumn}% Align table columns on decimal point
\usepackage{bm}% bold math
\usepackage{quantikz}
\usepackage{times}
\begin{document}

\title{Channel Attention for Quantum Convolutional Neural Networks}% Force line breaks with \\

\author{Gekko Budiutama}
\email{bgekko@quemix.com}
\affiliation{
Laboratory for Materials and Structures,
Institute of Innovative Research,
Tokyo Institute of Technology,
Yokohama 226-8503,
Japan
}
\affiliation{Quemix Inc.,
Taiyo Life Nihonbashi Building,
2-11-2,
Nihonbashi Chuo-ku, 
Tokyo 103-0027,
Japan}

\author{Shunsuke Daimon}
\email{daimon.shunsuke@qst.go.jp}
\affiliation{Quantum Materials and Applications Research Center, National Institutes for Quantum Science and Technology, Tokyo, 152-8550, Japan}

\author{Hirofumi Nishi}
\affiliation{
Laboratory for Materials and Structures,
Institute of Innovative Research,
Tokyo Institute of Technology,
Yokohama 226-8503,
Japan
}
\affiliation{Quemix Inc.,
Taiyo Life Nihonbashi Building,
2-11-2,
Nihonbashi Chuo-ku, 
Tokyo 103-0027,
Japan}

\author{Ryui Kaneko}
\affiliation{Waseda Research Institute for Science and Engineering, Waseda University, Shinjuku, Tokyo 169-8555, Japan}
\affiliation{Physics Division, Sophia University, Chiyoda, Tokyo 102-8554, Japan}

\author{Tomi Ohtsuki}
\affiliation{Physics Division, Sophia University, Chiyoda, Tokyo 102-8554, Japan}

\author{Yu-ichiro Matsushita}%
\affiliation{
Laboratory for Materials and Structures,
Institute of Innovative Research,
Tokyo Institute of Technology,
Yokohama 226-8503,
Japan
}
\affiliation{Quemix Inc.,
Taiyo Life Nihonbashi Building,
2-11-2,
Nihonbashi Chuo-ku, 
Tokyo 103-0027,
Japan}%
\affiliation{Quantum Materials and Applications Research Center, National Institutes for Quantum Science and Technology, Tokyo, 152-8550, Japan}%


\begin{abstract}
Quantum convolutional neural networks (QCNNs) have gathered attention as one of the most promising algorithms for quantum machine learning. Reduction in the cost of training as well as improvement in performance is required for practical implementation of these models. In this study, we propose a channel attention mechanism for QCNNs and show the effectiveness of this approach for quantum phase classification problems. Our attention mechanism creates multiple channels of output state based on measurement of quantum bits. This simple approach improves the performance of QCNNs and outperforms a conventional approach using feedforward neural networks as the additional post-processing.
\end{abstract}

\maketitle
Quantum neural networks have emerged as one of the promising approaches to perform machine learning tasks on noisy intermediate-scale quantum computers. Quantum neural networks employ a variational quantum ansatz, the parameters of which are iteratively optimized using classical optimization routines \cite{farhi2018classification, PhysRevA.98.032309, PhysRevA.101.032308, cerezo_variational_2021, PhysRevLett.131.140601, PhysRevResearch.1.033063}. Among various quantum neural network models, quantum convolutional neural networks (QCNNs) have been proposed as the quantum equivalent of the classical convolutional neural networks \cite{cong_quantum_2019, PhysRevA.102.012415}. Unlike other quantum neural networks \cite{mcclean_barren_2018, cerezo_cost_2021, PRXQuantum.2.040316, anschuetz_quantum_2022}, QCNNs do not exhibit the exponentially vanishing gradient problem (the barren plateau problem) \cite{pesah_absence_2021, zhao_analyzing_2021, cervero_martin_barren_2023}. This guarantees QCNNs' trainability under random initial parameterization. A recent study has also demonstrated the viability of this approach on real quantum computers \cite{herrmann_realizing_2022}. Due to these factors, QCNNs have gained significant attention in the field of quantum machine learning. They have demonstrated success in various learning tasks that involve quantum states as direct inputs, such as quantum phase classification \cite{herrmann_realizing_2022, liu_model-independent_2023, umeano2023learn, PhysRevB.107.L081105}. While having great promise, reduction in training costs is essential for QCNNs' practicality due to the current limitations and high expenses of today's quantum computer \cite{Perdomo-Ortiz_2018, Moll_2018, 8936946, wack2021quality, lubinski_application-oriented_2023, russo_testing_2023}. Furthermore, enhancing the performance of QCNNs is crucial for the efficacy of this approach.
To address these challenges, various hybrid QCNN-classical machine learning approaches have been reported. Here, the outputs of the QCNNs are used as the input to classical machine learning models. Fully connected feedforward neural networks are the most common classical models utilized in these approaches \cite{broughton2021tensorflow, sengupta_quantum_2021, sebastianelli_circuit-based_2022, li_image_2022, huang_image_2023}. However, the significant computational workload may become infeasible with the increasing complexity of tasks as well as QCNN models.

In this work, we propose a channel attention scheme for QCNNs. Inspired by classical machine learning literature \cite{bahdanau_neural_2016, xu_show_2016, kim_structured_2017, vaswani_attention_2017, ferrari_cbam_2018, bilan2018positionaware, Wang_2020_CVPR, Choi_Kim_Han_Xu_Lee_2020}, the proposed attention for QCNNs creates channels of output state by the measurement of qubits that are discarded in the conventional QCNN models. A weight parameter is then assigned to each channel indicating the difference in importance. Our results show that incorporating this straightforward step leads to a remarkable reduction in the error of the QCNN models for the task of quantum phase classification. This improvement was achieved without major alteration to the existing QCNN models and with only a minor increase in the number of parameters.

% \section{Definition of Channel Attention}
\begin{figure*}[ht]
    \centering
    \includegraphics[width=1\textwidth]{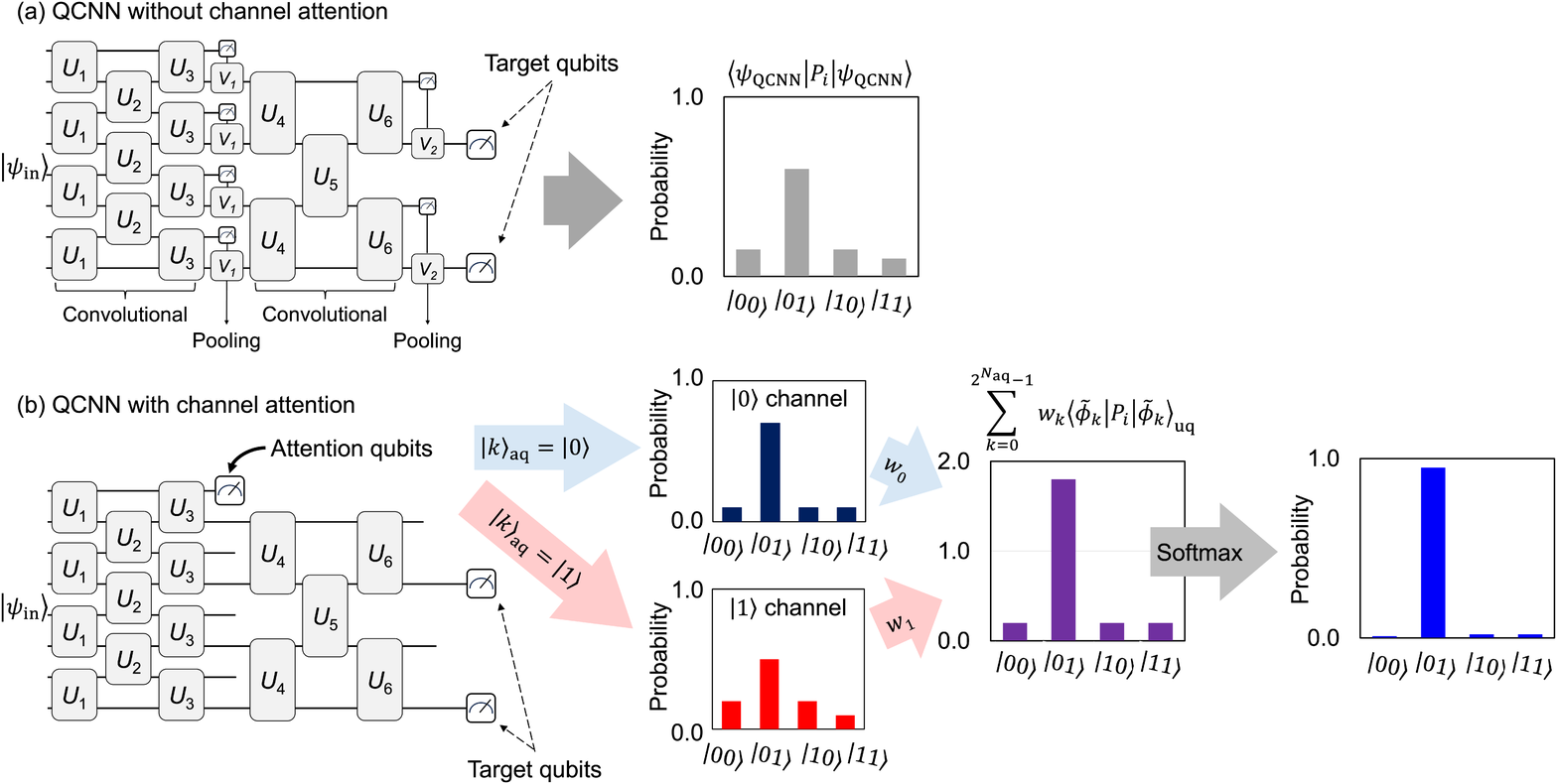}
    \caption{(a) A schematic illustration of the conventional QCNN. Here, $U_1$ to $U_6$ are parameterized gates that compose the convolutional layers. Controlled-$V_1$ and $V_2$ are controlled rotational gates that create the pooling layer. Here, $|\psi_{\mathrm{in}} \rangle$, $P_i$, $|\psi_{\mathrm{QCNN}} \rangle$ represent the input state, the projection operators for the $|i\rangle_{\mathrm{tq}}$ state, and the output state of the QCNN just before the measurement of the target qubits respectively. In the text, this model is referred to as QCNN without channel attention. (b) The proposed channel attention for QCNNs. Channels are created based on the measurement of attention qubits. The number of channels follows \(2^{N_{\mathrm{aq}}}\) where \(N_{\mathrm{aq}}\) is the number of attention qubits. In the figure above, a model with \(N_{\mathrm{aq}}=1\) and \(N_{\mathrm{tq}}=2\) is shown. Here, $|k\rangle_{\mathrm{aq}}$, $w_{k}$, $|\tilde{\phi}_k\rangle_{\mathrm{uq}}$ represent the state of attention qubits, weights of each channel, and the state of unmeasured qubits just before the measurement of the target qubits.
    }
    \label{fig1}
\end{figure*}
\begin{figure}[ht]
    \centering
    \includegraphics[width=0.45\textwidth]{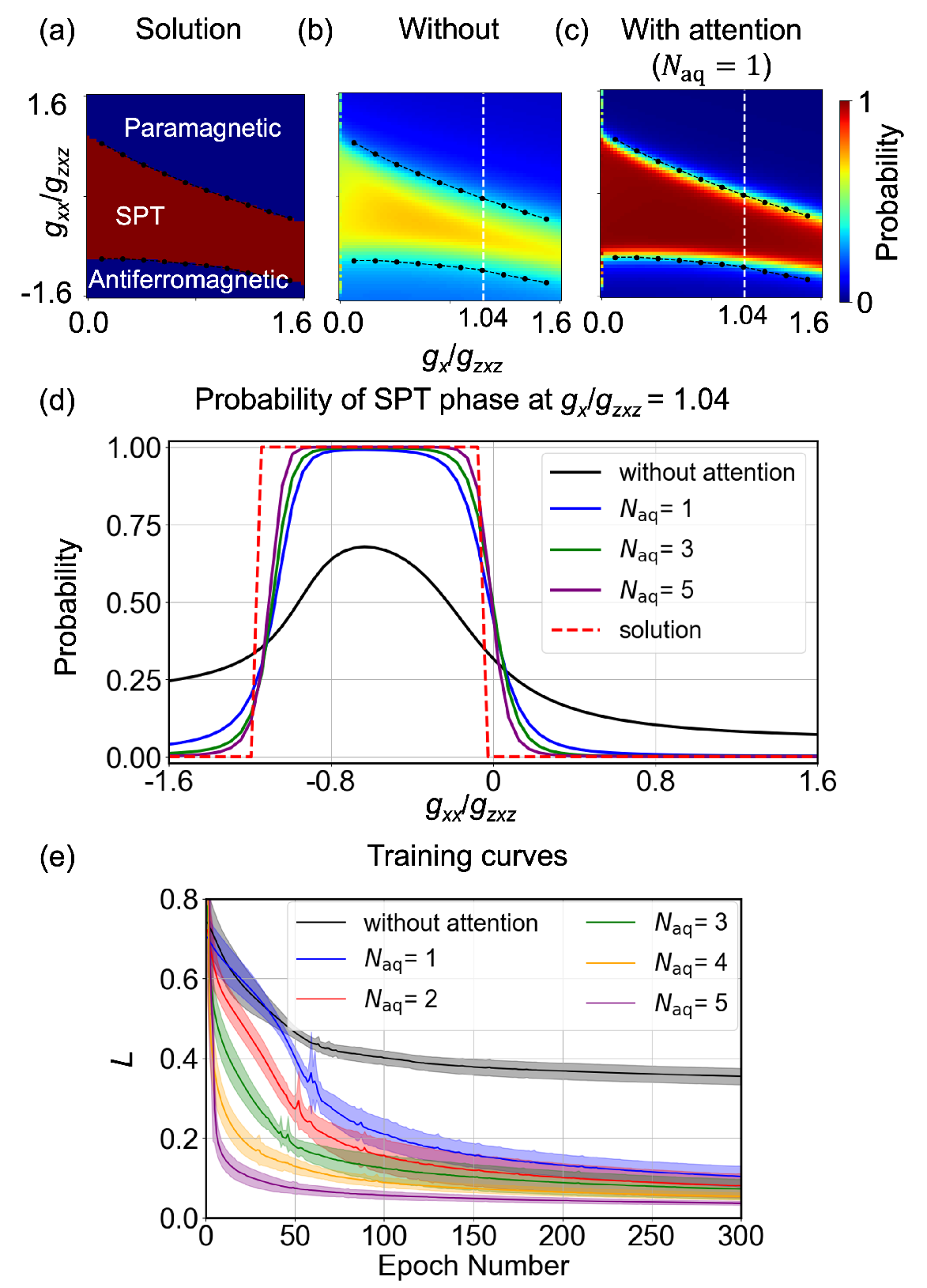}
    \caption{(a) The phase diagram of the Hamiltonian described in Eq. (6) as reported in literature \cite{cong_quantum_2019} estimated using classical method. The classification of the quantum phases using QCNN without (b) and with channel attention (c). Here, $g_x$, $g_{xx}$, and $g_{zxz}$ are the parameters of the Hamiltonian described in Eq. (6). Color gradation shows the probability of being in the SPT phase. (d) The probability of being in the SPT phase at $g_x/g_{zxz}=1.04$ of QCNN without (black) and with channel attention (color) at 300 epochs. (e) The training curves of QCNN without (black) and with channel attention (color). The shaded region represents the standard deviation for 10 random initial parameters.}
    \label{fig2}
\end{figure}
\begin{figure}[ht]
    \centering
    \includegraphics[width=0.45\textwidth]{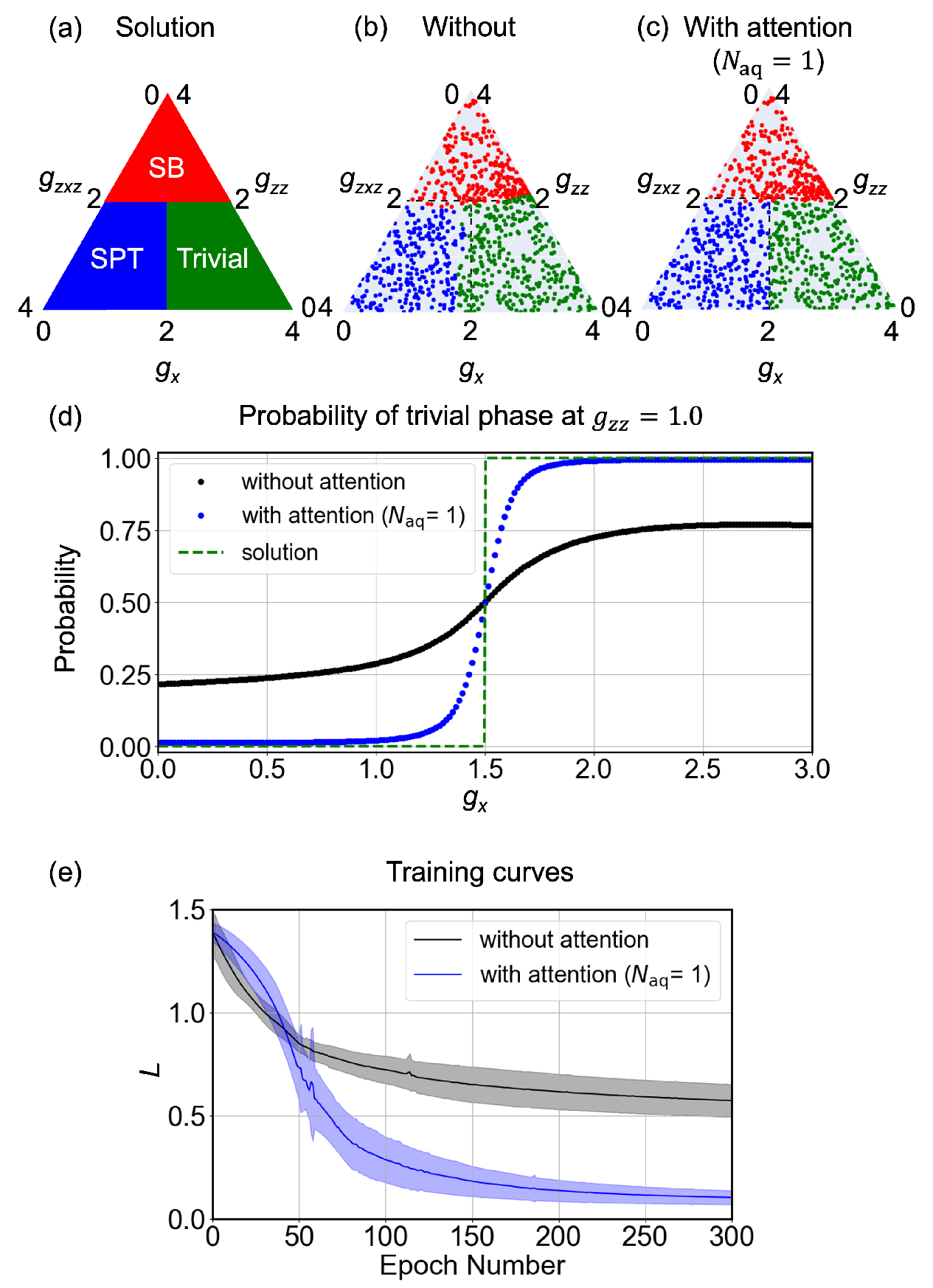}
    \caption{(a) The phase diagram of the Hamiltonian described in Eq. (7) as reported in literature \cite{PhysRevB.96.165124, PhysRevLett.120.057001, smith_crossing_2022}. The classification of the quantum phases using QCNN without (b) and with channel attention (c). Here, $g_{zxz}$, $g_{zz}$, and $g_{x}$ are the parameters of the Hamiltonian described in Eq. (7). (d) The probability of being in the trivial phase at $g_{zz}=1.0$ of QCNN without (black) and with channel attention (blue) at 300 epochs. (e) The training curves of QCNN without (black) and with channel attention (blue). The shaded region represents the standard deviation for 10 random initial parameters.}
    \label{fig3}
\end{figure}
\begin{figure}[ht]
    \centering
    \includegraphics[width=0.45\textwidth]{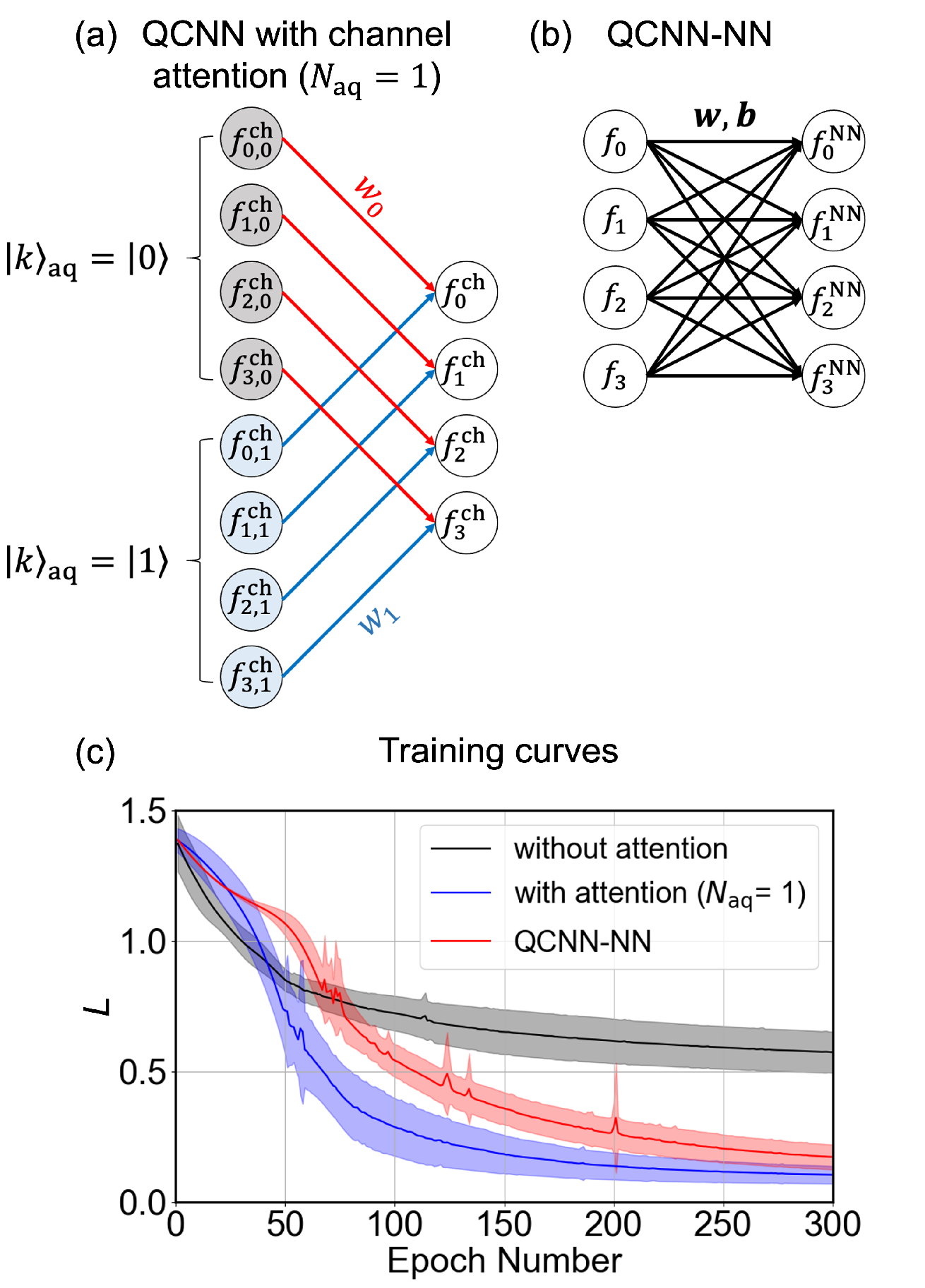}
    \caption{Illustrations of the post-processing method for QCNN with channel attention (a) and QCNN-NN (b). Here, $f_{i,k}^{\mathrm{ch}}$, $f_{i}^{\mathrm{ch}}$ represent the output probabilities from the QCNN under $k$-th channel and the final output probabilities from the QCNN with channel attention respectively. $w_0$ and $w_1$ are the attention parameters. $f_{i}$ and $f_{i}^{\mathrm{NN}}$ represent the output probabilities from the QCNN and the final output of the QCNN-NN model. $\textbf{\textit{w}}$ and $\textbf{\textit{b}}$ are the weight and bias parameters of the neural network. (c) The training curves of QCNN without (black) and with channel attention (blue), and QCNN-NN (red). The shaded region represents the standard deviation for 10 random initial parameters.}
    \label{fig4}
\end{figure}
As shown in Fig. 1 (a), conventional QCNNs are composed of convolutional and pooling layers. The convolutional layers consist of parallel local unitary gates that can be either uniformly or independently parameterized within a single layer \cite{cong_quantum_2019, liu_model-independent_2023} ($U_1$ to $U_6$ in Fig. 1 (a)). The pooling layers consist of controlled rotational gates that act on a single qubit (controlled-$V_1$ and $V_2$ in Fig. 1 (a)). Here, the control qubits on the pooling layer are discarded onwards, effectively reducing the number of qubits on the circuit. Following convolutional and pooling layers, some of the qubits are then measured as the final output of the model. From here on these qubits are called \emph{target qubits}.
% =========Additional from Nishi-san-Part1-START=========
Measuring these target qubits yields the prediction output:
\begin{gather}
    f_{i}(|\psi_{\mathrm{in}} \rangle; \boldsymbol{\theta}) 
    =
    \langle \psi_{\mathrm{QCNN}} |P_i| \psi_{\mathrm{QCNN}} \rangle,
\end{gather}
where $|\psi_{\mathrm{in}} \rangle$, $\boldsymbol{\theta}$, $|\psi_{\mathrm{QCNN}} \rangle$ represent the input state, the parameters of the QCNN, and the output state of the QCNN just before the measurement of the target qubits respectively.
In this study, the observables $\{P_i\}_{i=0,\ldots, 2^{N_\mathrm{tq}}-1}$ represent the projection operators for the $|i\rangle_{\mathrm{tq}}$ state, which act on the target qubits. $N_{\mathrm{tq}}$ is the number of target qubits.
% =========Additional from Nishi-san--Part1-END=========

The proposed QCNN with channel attention is realized by performing measurement(s) on the control qubits of the pooling layer (Fig. 1 (b)). From here on these qubits are called \emph{attention qubits}. The measurement result is used to create multiple channels of output state. Different weights are given to each channel representing their importance. The weighted outputs are summed and then subjected to a softmax function. The mathematical formalization of the attention for the QCNNs is shown below. 

% =========Additional from Nishi-san-Part2-START=========
Among various states of attention qubits, the entangled state collapses into a single state $|\phi_{k}\rangle_{\mathrm{uq}}$ upon measurement. Here, the measurement of attention qubits is represented as $|k\rangle_{\mathrm{aq}}$ and uq denotes unmeasured qubits in the pooling layer:
\begin{gather}
    \sum_{k=0}^{2^{N_{\mathrm{aq}}}-1}
    |\phi_k \rangle_{\mathrm{uq}} \otimes |k\rangle_{\mathrm{aq}}
    \to
    \frac{1}{\mathcal{N}_k} |\phi_k\rangle_{\mathrm{uq}} 
    \equiv |\tilde{\phi}_k\rangle_{\mathrm{uq}},
\end{gather}
where  
$   
    \mathcal{N}_k 
    = 
    \left \|
        |\phi_k\rangle_{\mathrm{uq}} 
    \right\|
$ 
is the normalization constant.
% =========Additional from Nishi-san-Part2-END=========
% =========Additional from Nishi-san-Part3-START=========
The channels of states are determined by measuring the target qubits based on the states of the attention qubits:
\begin{gather}
    f_{i,k}^{\mathrm{ch}} (|\psi_{\mathrm{in}} \rangle; \boldsymbol{\theta}) 
    =
    \langle \tilde{\phi}_k |P_i| \tilde{\phi}_k\rangle_{\mathrm{uq}}.
\end{gather}
Different weights $w_k \in \mathbb{R}$ are given to each channel, representing their importance. These weights are optimized during the training of the model. The final prediction data of the QCNN with channel attention is normalized using the softmax function:
\begin{gather}
    f_{i}^{\mathrm{ch}}(|\psi_{\mathrm{in}}\rangle; \boldsymbol{\theta}, \boldsymbol{w})
    =
    \operatorname{softmax} \left(
        \sum_{k=0}^{2^{N_{\mathrm{aq}}}-1}
        w_{k}
        f_{i,k}^{\mathrm{ch}} 
        (|\psi_{\mathrm{in}} \rangle; \boldsymbol{\theta})
    \right),
\end{gather}
where the softmax function is defined as
$
    \operatorname{softmax}\left(x_i\right)
    =
    e^{x_i}/\sum_{j=0}^{2^{N_{\mathrm{tq}}}-1} e^{x_j}
$.

% Need editing from here
For classification, QCNNs map an input $|\psi_{\mathrm{in}}\rangle$ to a label $\boldsymbol{y}$, where $\boldsymbol{y}$ denotes a one-dimensional one-hot binary vector of length $2^{N_{\mathrm{tq}}}$.
An input state is classified into $i$th category when $y_i=1$.
For a training dataset of size $M$, denoted as
$
    \{(
        |\psi_{\mathrm{in}}^{\alpha}\rangle,
        \boldsymbol{y}^{\alpha}
    )\}_{\alpha=1,\ldots,M}
$, the parameters $\boldsymbol{\theta}$ of the QCNN are determined by minimizing a loss function. In this study, the following log-loss function is utilized:
\begin{gather}
    L
    =
    - \frac{1}{M}
    \sum_{\alpha=1}^{M}
    \sum_{i=0}^{2^{N_{\mathrm{tq}}} - 1}
    y_i^{\alpha}
    \log f_{i}(|\psi_{\mathrm{in}}^{\alpha}\rangle; \boldsymbol{\theta}).
\end{gather}

% =========Additional from Nishi-san-Part3-END=========
% \section{The effect of attention on quantum phases classifications}

The effect of attention on the performance of QCNNs was investigated in the task of classification of quantum phases. Firstly, we utilized QCNNs to classify quantum phases in 1D many-body systems of spin-1/2 chains with open boundary conditions defined by the following Hamiltonian:
\begin{equation}
 H = -g_{zxz} \sum_{i=1}^{N-2} Z_iX_{i+1}Z_{i+2}-g_x \sum_{i=1}^{N}X_i-g_{xx} \sum_{i=1}^{N-1}X_iX_{i+1}
\label{eq:energy},
\end{equation}
where $Z_i$ ($X_i$) is the Pauli Z (X) operator for the spin at site $i$ and $g_{zxz}$, $g_x$, and $g_{xx}$ are the parameters of the Hamiltonian. Based on the given parameters, the ground state of the system can manifest as a symmetry-protected topological (SPT), antiferromagnetic, or paramagnetic phase (Fig. 2 (a)) \cite{cong_quantum_2019}.

The task is to determine whether a given ground state belongs to the SPT phase or not, and thus a binary classification problem. The training dataset was composed of 40 equally spaced points with $g_{xx}/g_{zxz} \in [-1.6, 1.6]$ along the line \(g_x/g_{zxz}=0.8\). The test dataset was composed of 4096 data points with ($g_x/g_{zxz} \in [0, 1.6]$ and $g_{xx}/g_{zxz} \in [-1.6, 1.6]$). Here, a system with 9 spins was considered. QCNN without and with channel attention were used to perform this classification. Here, the SPT phase is assigned to $\ket{0}$ and the non-SPT phase is assigned to $\ket{1}$. Details regarding the circuit model and numerical simulation of the QCNNs were described in Supplemental Material S1 \cite{SM}.

Figures 2 (b) and (c) show the classification results for QCNN without and with channel attention ($N_{\mathrm{aq}}=1$) after 300 epochs of training. The ability to identify the SPT and non-SPT regions is observed in both QCNN models. However, QCNN with channel attention improved upon this result by emphasizing the probability of the correct state occurring. Figure 2 (d) shows the probability of data points being classified as SPT phase at $g_{xx}/g_{zxz} \in [-1.6, 1.6]$ and fixed $g_x/g_{zxz}=1.04$. Here, it is observed that channel attention enhanced the predictive ability of the QCNN, bringing it into closer alignment with the reported solution. Increasing the $N_{\mathrm{aq}}$ further refines the accuracy of the model to identify the correct phase. This improvement is observed especially around the phase boundaries (Supplemental Material Fig. S2) \cite{SM}. Figure 2 (e) shows the learning curves of QCNN models without (black) and with (color) attention. For all $N_{\mathrm{aq}}$, QCNN with channel attention shows significantly lower $L$ for the training dataset, approximately 3 ($N_{\mathrm{aq}}$=1) to 10 ($N_{\mathrm{aq}}$=5) times lower compared to the original QCNN. Increasing $N_{\mathrm{aq}}$ from 1 to 5 further reduces the final $L$ for the training dataset and increases the rate of convergence. Moreover, QCNN with channel attention maintained the generality of the QCNN for quantum phase classification problems, as correct classification over large number of dataset with extensive parameter range (4096 data points, $g_x/g_{zxz} \in [0, 1.6]$ and $g_{xx}/g_{zxz} \in [-1.6, 1.6]$) was achieved from a few training data with limited parameter range (40 data points, $g_{xx}/g_{zxz} \in [-1.6, 1.6]$ along the line \(g_x/g_{zxz}=0.8\)).

The performance of QCNNs with and without channel attention was also compared on an open boundary system with \(\mathbb{Z}_2 \times \mathbb{Z}_2^{T}\) symmetric Hamiltonian:
\begin{equation}
    \label{eq:energy_2}
    H = g_{zxz} \sum_{i=2}^{N-1} Z_{i-1}X_iZ_{i+1}-g_x \sum_{i=1}^{N}X_i-g_{zz} \sum_{i=1}^{N-1}Z_{i}Z_{i+1},
\end{equation}
where $g_{zxz}$, $g_x$, and $g_{zz}$ are the parameters of the Hamiltonian. 
Based on the given parameters, the ground state of the system can manifest as a symmetry-protected topological phase with cluster state (SPT), trivial phase, or symmetry-broken phase (SB), and thus a ternary classification problem. Figure 3 (a) illustrates the correct classification for the given Hamiltonian (Eq. (7)) using a ternary graph under $g_{zxz}+g_x+g_{zz}=4$ \cite{PhysRevB.96.165124, PhysRevLett.120.057001, smith_crossing_2022}. Under this condition, a dataset comprising 900 evenly distributed data points, with 300 data points from each phase, was employed for the investigation with a training-to-testing ratio of 70:30. Here, a system with 9 spins was considered. The classification was performed by assigning each class to a specific state of the two target qubits ($\ket{00}$ for trivial, $\ket{01}$ for SB, and $\ket{10}$ for SPT). 

Classification results of QCNN without and with channel attention are shown in Fig. 3 (b) and (c) respectively. Here, it is observed that the QCNN without channel attention was not able to correctly classify the phases near the boundary, which is particularly noticeable at the boundary between the SPT and trivial phases. In contrast, the QCNN with channel attention returned a significantly better classification for all three phases. Figure 3 (d) shows the probability of data points being classified as the trivial phase at $g_{x} \in [0, 3]$, fixed $g_{zz}=1.0$, and $g_{zxz}+g_x+g_{zz}=4$. Here, it is observed that QCNN with channel attention improved upon the prediction output of the QCNN, bringing the prediction output closer to the reported solution.
Figure 3 (e) shows the training curves of QCNN without (black) and with channel attention (blue) for the ternary classification problem. Here, our attention method reduces the $L$ for the training dataset by approximately 6 times and increases the classification confidence by emphasizing the correct output state probability to close to 1 (Supplemental Material Fig. S3) \cite{SM}.

% \section{Channel Attention vs. All-connected Feedforward Neural Network}
The proposed channel attention acts as a classical post-processing protocol for the QCNNs (Fig. 4 (a)). As mentioned previously, the most common QCNN-classical machine learning approaches use fully connected neural networks for classical post-processing \cite{broughton2021tensorflow, sengupta_quantum_2021, sebastianelli_circuit-based_2022, li_image_2022, huang_image_2023} (from here on, called QCNN-NN). Here, the performance of QCNN with channel attention is directly compared with the QCNN-NN for the ternary classification problem. 

The model used for the comparison is a fully connected feedforward with four nodes on the input layer and four nodes on the output layer with no hidden layer (Fig. 4 (b)). This NN utilized 20 parameters (16 weights, 4 biases) and a softmax activation function. 
The final prediction of this QCNN-NN model is defined as:
{\small
\begin{gather}
    f_{i}^{\mathrm{NN}}(|\psi_{\mathrm{in}}\rangle; \boldsymbol{\theta}, \boldsymbol{w}, \boldsymbol{b})
    =
    \operatorname{softmax} \left(b_i +
        \sum_{k=0}^{2^{N_{\mathrm{tq}}}-1}
        w_{i,k}
        f_{k}
        (|\psi_{\mathrm{in}} \rangle; \boldsymbol{\theta})
    \right),
\end{gather}} where $\boldsymbol{w}$ and $\boldsymbol{b}$ are the weight and bias parameters of the neural network. 

Figure 4 (c) shows the training curves of the QCNN with channel attention (blue) and QCNN-NN (red) for the ternary classification problem. 
It is observed here that QCNN with channel attention converged faster and achieved lower $L$ for the training dataset when compared to the QCNN-NN.
Furthermore, with only 2 additional parameters ($w_0$ and $w_1$), QCNN with channel attention outperforms QCNN-NN (20 additional parameters) with the average $L$ for the testing dataset of $9.9 \times 10^{-2}$ and $16.5 \times 10^{-2}$ respectively at 300 training epochs (Table I). 
\begin{table}[ht]
  \centering
  \caption{The average $L$ for the testing dataset (ternary problem) of the QCNNs at 300 epochs.}
  \resizebox{8.6 cm}{!}{
  \begin{tabular}{ccc}
    \hline
    QCNN & QCNN with channel attention & QCNN-NN \\
         & \scriptsize (2 additional parameters) & \scriptsize (20 additional parameters) \\
    \hline
    $0.555 \pm 0.062$ & $0.099 \pm 0.033$ & $0.165 \pm 0.044$ \\
    \hline
  \end{tabular}
  }
\end{table}
This shows that QCNN with channel attention is a more effective hybrid approach compared to the QCNN-NN as it utilizes a significantly lower number of parameters to achieve better performance.

In conclusion, we introduced a channel attention mechanism for the quantum convolutional neural networks (QCNNs). In our approach, channels of output state are created by additional measurements of qubit(s) that are discarded in the conventional QCNN models. The importance of each channel is then computed. Our results showed that these straightforward steps led to a significant increase in the performance of the QCNNs without any major alteration to the already existing models. Integrating attention to QCNN models reduced the classification error of quantum phase recognition problems by at least 3 fold for the binary classification problem and 6 fold for the ternary classification problem. Comparison between the QCNN with channel attention with QCNN-NN showed that QCNN with channel attention outperforms QCNN-NN with significantly fewer parameters. Thus, the proposed method is an effective and low-cost method that substantially improves the practicality, versatility, and usability of the QCNNs.

\begin{acknowledgments}
This work was supported by JSPS KAKENHI under Grant-in-Aid for Scientific Research No. 21H04553, No. 20H00340, and No. 22H01517, JSPS KAKENHI under Grant-in-Aid for Transformative Research Areas No. JP22H05114, JSPS KAKENHI under Grant-in-Aid for Early-Career Scientists No. JP21K13855, and by JST Grant Number JPMJPF2221. This study was partially carried out using the TSUBAME3.0 supercomputer at the Tokyo Institute of Technology and the facilities of the Supercomputer Center, the Institute for Solid State Physics, the University of Tokyo. The author acknowledges the contributions and discussions provided by the members of Quemix Inc.
\end{acknowledgments}
\bibliography{manuscript}% Produces the bibliography via BibTeX.

\end{document}